# n-Type diamond synthesized with *tert*-butylphosphine for long spin coherence times of perfectly aligned NV centers

Riku Kawase,[a] Hiroyuki Kawashima,[a] Hiromitsu Kato,[b] Norio Tokuda,[c] Satoshi Yamasaki,[c] Masahiko Ogura,[b] Toshiharu Makino[b], and Norikazu Mizuochi[a,d]

[a] *Institute for Chemical Research, Kyoto University, Gokasho, Uji, Kyoto 611-0011, Japan*

[b] *National Institute of Advanced Industrial Science and Technology (AIST), Tsukuba, Ibaraki 305-8568, Japan*

[c] *Graduate School of Natural Science and Technology, Kanazawa University, Kanazawa, Ishikawa 920-1192, Japan*

[d] *Center for Spintronics Research Network, Kyoto University, Uji, Kyoto 611-0011, Japan*

## Abstract

The longest spin coherence times for nitrogen-vacancy (NV) centers at room temperature have been achieved in phosphorus-doped *n*-type diamond. However, difficulty controlling impurity incorporation and the utilization of highly toxic phosphine gas in the chemical vapor deposition (CVD) technique pose problems for the growth of *n*-type diamond. In the present study, *n*-type diamond samples were synthesized by CVD using *tert*-butylphosphine, which is much less toxic than phosphine. The unintentional incorporation of nitrogen was found to be suppressed by incrementally increasing the gas flow rates of $H_2$ and $CH_4$. Hall measurements confirmed *n*-type conduction in three measured samples prepared under different growth conditions. The highest measured Hall mobility at room temperature was 422 $cm^2/(Vs)$. In the sample with the lowest nitrogen concentration, the spin coherence time ($T_2$) increased to 1.62 ± 0.10 ms. Optically detected magnetic resonance spectra indicated that all of the measured NV centers were aligned along the [111] direction. This study provides appropriate CVD conditions for growing phosphorus-doped *n*-type diamond with perfectly aligned NV centers exhibiting long spin coherence times, which is important for the production of quantum diamond devices.

## I. INTRODUCTION

In quantum science and technology, long coherence times are required to enhance the sensitivity of quantum sensors and prolong the quantum memory time of quantum information devices.[1,2,3,4] Among the candidate solid-state materials for use in quantum devices, diamonds with nitrogen-vacancy (NV) centers have long coherence times[3,5,6] even at room temperature.[7,8,9] Over the past few decades, coherence times have increased with advances in diamond growth technology; impurities and $^{13}$C nuclear spins that are noise sources have been removed.[7,9,10,11] By exploiting the excellent characteristics of diamonds with NV centers, researchers have demonstrated their suitability for use in quantum sensing.[2,3,4,12,13,14]

In recent years, long coherence times have been observed for NV centers in phosphorus-doped *n*-type diamonds synthesized by chemical vapor deposition (CVD), which are the longest coherence times reported among solid-state electron spin materials at room temperature.[9] This is important from the viewpoint of stabilizing the charge state of NV centers[15,16] and applying *n*-type diamond in quantum devices with semiconductor characteristics. From the perspective of quantum sensors, CVD growth also has the advantage of aligning the NV axes.[17,18,19] Recently, the use of the CVD growth method to create NV centers has been studied.[10,20,21,22,23] Because of the difficulty of controlling the incorporation of impurities during CVD, growth of *n*-type diamond has been limited.[24,25,26,27] In addition, *n*-type diamond is usually synthesized using phosphine gas,[24] which is highly toxic. However, *tert*-butylphosphine (TBP) has a safety advantage over phosphine because its toxicity level is several orders of magnitude lower.[28]

In the present paper, phosphorus-doped diamond was synthesized using TBP and its *n*-type conductivity was confirmed from Hall measurements. The *n*-type diamond was synthesized in a CVD growth chamber, which can potentially be used to synthesize materials at a high growth rate and with low power consumption.[29,30] We investigate the impurity concentration, coherence time ($T_2$), and the NV axis alignment dependence on the growth conditions.

## II. EXPERIMENTAL

The diamond samples were grown on Ib-type (111) diamond substrates (grown via the high-pressure–high-temperature method, dimensions 2 × 2 × 0.5 mm$^3$ or 2 × 2 × 0.3 mm$^3$) by microwave plasma-assisted chemical vapor deposition (MPCVD). The main source gas was $^{12}$C-enriched $CH_4$ (purity: greater than 8N, [$^{13}$C] < 0.005 at%) or natural abundant $CH_4$ (purity: 6N) (used to grow the samples for secondary-ion mass spectrometry (SIMS) measurements) and TBP (purity: 3N3) diluted with $H_2$ (purity: greater than 8N).[31] Table 1 shows the growth conditions and the properties of the resultant films, including the nitrogen and phosphorus concentrations and the $T_2$ of the samples grown under the four growth conditions labeled O, A, B, and C. For each growth type, the total gas flow rate and $CH_4$ gas flow rate were different: 100 sccm, 0.1 sccm ($CH_4/H_2$: 0.1%) for type O; 400 sccm, 0.4 sccm ($CH_4/H_2$: 0.1%) for type A; 100 sccm, 1.0 sccm ($CH_4/H_2$: 1.0%) for type B; and 400 sccm, 1.0 sccm ($CH_4/H_2$: 0.25%) for type C. The TBP flow rate was adjusted to make the phosphorus concentration $10^{16}$–$10^{17}$ cm$^{-3}$ in the film, corresponding to that of the sample with the longest reported $T_2$ at room temperature (~2.4 ms).[9] Regarding growth types A and B, a sufficiently high phosphorus concentration was achieved by incorporating residual phosphorus in the CVD chamber. The gas pressure was 25 kPa in common. The off-angle of the substrates corresponded to 1°–3° along the $[\bar{1}\bar{1}2]$ direction. After the growth, samples were cleaned with mixed acid ($H_2SO_4$ : $HNO_3$ = 3:1) at 150°C for 30 min. The elemental composition of the CVD layers was analyzed by SIMS. The growth rates estimated from the SIMS results were between 1 and 6 μm/h.

For Hall measurements, heavily phosphorus-doped layers were deposited at the electrode using a selective growth method to reduce the contact resistance.[27,29] The electrode fabrication method was as follows. First, Ti was deposited onto the entire surface (except for the electrode area) of phosphorus-doped diamond layers grown by the MW plasma CVD method. Photolithography and lift-off methods were used to form the Ti mask pattern.[32] With the Ti mask in place, heavily phosphorus-doped diamond layers (concentration on the order of $10^{20}$ cm$^{-3}$) were grown by MW plasma CVD. In this way, heavily phosphorus-doped films were grown only in the area without the Ti mask. After the Ti mask was removed by acid washing, Ti (30 nm)/Pt (30 nm)/Au (100 nm) films were deposited by electron beam evaporation onto the area with heavily phosphorus-doped diamond films. Finally, the samples were annealed in a muffle furnace at 420°C for 30 min in air. This process further suppressed the contact resistance by hybridizing the interface between the metal and diamond to form TiC.

The photoluminescence (PL) from the NV centers was measured at room temperature using a home-built confocal microscope system. The NV centers were excited with a 532 nm laser, and the PL was detected with avalanche photodiode detectors and a spectrometer. Microwaves were generated by a thin copper wire fabricated on the surface of the sample to make the optically detected magnetic resonance (ODMR) measurement.

## III. RESULTS AND DISCUSSION

Figure 1a shows a PL image acquired by confocal scanning microscopy. The PL spectrum of the bright spot, as recorded using a spectrometer, is shown in Figure 1b. The results indicate that the bright spots were the PL of the NV centers. At a bright spot, the normalized second-order autocorrelation function $g^{(2)}(\tau)$ was measured (Figure 1c). The value of $g^{(2)}(0)$ was less than 0.5, indicating that the PL was emitted from a single NV center.

To investigate the orientation of the NV axis, a continuous wave (CW)-ODMR spectrum was acquired at room temperature (Figure 2a). A static magnetic field ($B \sim 59\ G$) was applied with a permanent magnet from the direction perpendicular to the diamond (111) surface, and microwave fields were applied to control the electron spin state ($|M_s\rangle = |0\rangle \Leftrightarrow |\pm1\rangle$) of the NV centers. The spin state could be read out optically from the decrease in the PL intensity of the NV centers at the magnetic

resonance frequency.

The ODMR spectrum was simulated on the basis of the following Hamiltonian:

$$\mathbf{H} = g_e \beta_e \tilde{\boldsymbol{S}} \cdot \boldsymbol{B} + \tilde{\boldsymbol{S}} \cdot \boldsymbol{D} \cdot \boldsymbol{S} \quad (1)$$

where an electron of spin $S = 1$ is considered and $\beta_e$ is the Bohr magneton.[17] Previously reported values for the zero-field splitting parameter ($D$ = 2.872 GHz) and the isotropic electron Zeeman $g$-value ($g_e = 2.0028$) were used.[33] Regarding the NV center, the ODMR can be explained mainly on the basis of Zeeman splitting ($g_e \beta_e \tilde{\boldsymbol{S}} \cdot \boldsymbol{B}$) and zero-field splitting ($\tilde{\boldsymbol{S}} \cdot \boldsymbol{D} \cdot \boldsymbol{S}$); thus, hyperfine splitting was neglected here.[34] In addition, we set the zero-field splitting parameter $E$ to 0 because the NV center exhibits $C_{3v}$ point symmetry.[34]

Figures 2b and 2c show the simulated spectra based on Eq. (1), where the direction of the magnetic field was assumed to be 0° (aligned with the [111] direction) and 109.47° (aligned with the $[1\bar{1}\bar{1}], [\bar{1}1\bar{1}]$, or $[\bar{1}\bar{1}1]$ direction). Because the measured resonance frequencies of the Zeeman splitting ($g_e \beta_e B_{[111]} \sim 320$ MHz) well correspond to those of the upper simulated result in Figures 2b and 2c, we can conclude that the NV axis is oriented in the [111] direction. ODMR measurements were performed for more than 100 NV centers of growth type B and type C samples to reveal the proportion of the NV axis orientation. Figure 2d shows the stacked ODMR spectra, which show the orientations of the NV axis for all the measured NV centers aligned in the [111] direction.

To obtain the Rabi frequency, Rabi oscillation measurements were conducted by changing the pulse length of the resonant microwaves. The $\pi$-pulse length was calculated from the Rabi frequency, and a Hahn echo pulse sequence ($\pi/2, \pi, \pi/2$-pulse) was applied to estimate $T_2$. Figure 3a shows the results of the Rabi oscillation measurements. The $\pi$-pulse length can be estimated from the sinusoidal fitting results. The typical Rabi contrast in our samples was 30–40%, which is approximately the same or slightly higher than that for intrinsic diamond.[9] We subsequently conducted Hahn echo measurements to evaluate $T_2$. Figure 3b shows the measurement scheme; a $\pi/2$-pulse was used as the final pulse to obtain the $|0\rangle$ state, and a $-\pi/2$-pulse (or, alternatively, a 3/2-pulse) was used as the final pulse to obtain the $|1\rangle$ state. Figure 3c shows two types of readout results ($S_{|0\rangle}, S_{|1\rangle}$), which were obtained by observing the $|0\rangle$ state and the $|1\rangle$ state, respectively. To remove common-mode noise, these two results were subtracted and normalized as $((S_{|0\rangle} - S_{|1\rangle})/(S_{|0\rangle} + S_{|1\rangle}))$.[35] The value of $T_2$ was estimated by exponential fitting ($\exp(-(\tau/T_2)^n)$) of the subtraction result. Figure 3d shows the results of the Hahn echo measurement from which the longest $T_2 = 1.62 \pm 0.10$ ms was obtained.

Figure 4a shows the measured $T_2$ for each growth type. The longest $T_2$ for growth-types O, A, B, and C was 0.38 ± 0.03 ms, 0.90 ± 0.06 ms, 1.38 ± 0.08 ms, and 1.62 ± 0.10 ms, respectively. Before we synthesized the growth-type A–C samples, we synthesized the growth-type O sample by setting the $H_2$ and $CH_4$ flow rates at 100 sccm and 0.1 sccm, respectively. For the growth-type O sample, the longest $T_2$ was 380 ± 30 μs (Table 1 and Figure 4a). When the $CH_4$ and the $H_2$ flow rates were increased for preparing the growth-type A–C samples, $T_2$ was increased (Figure 4a).

Figure 4b shows the nitrogen concentration in the CVD films with respect to the $CH_4$ and $H_2$ flow rates. These results indicate that the nitrogen concentration in the films was affected most strongly by increasing the gas flow rates. For each growth type, the nitrogen concentration decreased with increasing gas flow rate. In a previous study, the contribution of the substitutional nitrogen (P1 center) to $T_2$ was estimated to be 160 ± 12 μs ppm.[3] Given this relationship, we can infer whether the observed $T_2$ is limited by noise due to nitrogen impurities. In Figure 4a, the $T_2$ estimated from the nitrogen concentrations with a reported relationship 160 ± 12 μs ppm[3] are shown as green crosses to the right of the measured $T_2$. In this estimation, the nitrogen concentration measured by SIMS (Table 1 and Figure 4b) was used. Comparing the estimated $T_2$ of each growth type reveals that the longest or average $T_2$ of each growth type is approximately equal to the observed $T_2$. From this comparison, we concluded that the $T_2$ of each growth type is mainly dominated by the noise from nitrogen impurities. The comparison also reveals that the longest $T_2$ in the present study was shorter than the longest $T_2$ of 2.4 ms reported previously[9] because of the high concentration of nitrogen impurities. Notably, our estimated nitrogen concentration obtained by the SIMS measurement includes impurities other than P1 centers; however, nitrogen impurities other than P1 centers, such as NVH[36] or other complexes,[37] are often paramagnetic; they are therefore considered to have the same effect as a P1 center. Regarding the scatter of $T_2$ of each NV center in Figure 4a, the reasons are considered to be the different local surroundings of each one, such as paramagnetic defects and/or impurities. Next, we discuss the phosphorus concentration in each type.

Regarding the phosphorus concentration, $T_2$ has been reported to become longest at a phosphorus concentration of ~$6 \times 10^{16}$ $cm^{-3}$.[9] The $T_2$ for the TBP flow rate in the growth-type B and C samples are shown in Figures 5a and 5b, respectively. The phosphorus concentrations estimated from the SIMS measurements are also shown. Regarding the growth-type B sample, the phosphorus concentration was $4 \times 10^{16}$ $cm^{-3}$ with a TBP flow rate of 0 sccm. Regarding the growth-type C sample, the phosphorus concentration was $2 \times 10^{17}$ $cm^{-3}$ with a TBP flow rate of 1060 sccm. When we increased the flow rates, the $T_2$ decreased. A comparison between the TBP flow rate and the measured phosphorus concentration reveals that $T_2$ became shorter as the phosphorus concentration was increased beyond the reported optimum concentration ($6 \times 10^{16}$ $cm^{-3}$), which is consistent with the results of the previous study.[9] When combined with the effect of nitrogen impurities, the results show that the nitrogen concentration limits the longest $T_2$ and that the phosphorus concentration also affects $T_2$ when it increases beyond the optimum phosphorus concentration.

We investigated the electrical properties of the three phosphorus-doped diamond samples and confirmed their *n*-type conductivity by Hall effect measurements. Hall effect measurements are the most popular analytical approach to investigating the electrical properties of semiconductors. Four-terminal electrodes based on the van der Pauw method were used in the measurements.[38] The applied magnetic field was 0.5 T, and the temperature range was 300–900 K. However, for *n*-type phosphorus-doped diamond with a phosphorus concentration of ~$2 \times 10^{17}$ cm$^{-3}$, the contact resistance at the metal–diamond interface poses a problem for Hall effect measurements. The Fermi level at the metal–diamond interface is strongly pinned at ~4.3 eV below the bottom of the conduction band.[39] This causes the electrode to become a Schottky contact, which interferes with Hall effect measurements. The only solution is to increase the donor concentration and reduce the resistivity. Therefore, heavily phosphorus-doped layers were selectively grown under the electrodes to reduce the contact resistance.[27,29,32]

Figure 6a shows the temperature dependence of the carrier concentration, as obtained from the Hall measurement results, for three diamond samples. An *n*-type contribution is confirmed for all of the phosphorus-doped diamond films. With respect to the temperature dependence, the carrier concentration decreases exponentially as a function of $1/T$, indicating that band conduction, not hopping conduction, is dominant. This result is reasonable given the phosphorus concentration of $10^{16}$–$10^{17}$ cm$^{-3}$ in the diamond films, as determined by SIMS analysis.

Each dashed line in Figure 6a was fitted with the following carrier concentration equation for the measurement points of both samples:

$$\frac{n(N_{\mathrm{A}}+n)}{N_{\mathrm{D}}-N_{\mathrm{A}}-n}=\frac{N_{\mathrm{C}}}{\mathrm{g}}\exp\left(-\frac{E_{\mathrm{D}}}{k_{\mathrm{B}}T}\right) \quad (1)$$

where $n$ is the carrier concentration in the diamond film, $N_{\mathrm{D}}$ is the donor concentration, $N_{\mathrm{A}}$ is the acceptor concentration, and $E_{\mathrm{D}}$ is the ionization energy of the donor level. The acceptor concentration is unintentional, caused by impurities or defects. $k_{\mathrm{B}}$ is Boltzmann's constant, $T$ is the temperature, and $g$ is the degeneracy of the donor level (in the case of diamond, $g = 2$). $N_{\mathrm{C}}$ is the effective density of the conduction-band levels, expressed as a function of $T$ as

$$N_{\mathrm{C}}=2M_{\mathrm{C}}\left(\frac{2\pi m_{\mathrm{dos}}^{*}k_{\mathrm{B}}T}{h^{2}}\right)^{3/2} \quad (2)$$

where $h$ is Planck's constant and $M_{\mathrm{C}}$ is the number of equivalent minima in the conduction band (for diamond, $M_{\mathrm{C}} = 6$). $m_{\mathrm{dos}}^{*}$ is the density-of-state effective mass of electrons and is given by $m_{\mathrm{dos}}^{*} = (m_{\mathrm{l}}^{*}m_{\mathrm{t}}^{*}m_{\mathrm{t}}^{*})^{1/3}$, where $m_{\mathrm{l}}^{*}$ and $m_{\mathrm{t}}^{*}$ are the longitudinal and transverse effective mass of electrons, respectively. In this case, $m_{\mathrm{l}}^{*} = 1.4m_0$ and $m_{\mathrm{t}}^{*} = 0.36m_0$ were assumed, where $m_0$ is the free electron mass.[40]

Table 2 shows the parameters for each phosphorus-doped diamond film, as derived from Hall effect measurements and the aforementioned fitting calculations. The $E_{\mathrm{D}}$ values estimated from the fitting results were 0.53 eV for samples B1 and B2 and 0.62 eV for sample A2; these values are close to the standard value of 0.6 eV for phosphorus-doped diamond. The carrier compensation ratio ($\eta = N_{\mathrm{A}}/N_{\mathrm{D}}$) was estimated to be ~55% for sample B1 and B2 and ~64% for sample A2.

Figure 6b shows the temperature dependence of the electron mobility. The mobility decreases with increasing temperature for both samples and decreases due to acoustic phonon scattering ($\mu \propto T^{-3/2}$). In this temperature range, acoustic phonon scattering dominates the electron mobility. The temperature dependence changes at higher temperatures, which suggests that the electron mobility was dominated by intervalley phonon scattering, which is less sensitive to impurities. These results are in good agreement with the results of previous reports of phosphorus-doped diamond growth.[41,42]

The maximum mobility in this measurement was 422 $cm^2/(Vs)$ at room temperature for sample B1, 272 $cm^2/(Vs)$ at room temperature for sample B2, and 374 $cm^2/(Vs)$ at 340 K for sample A2. However, these values are lower than previously reported values for diamond films with the same phosphorus concentration.[43] Because of the high carrier compensation ratio, improving the crystal quality is considered necessary to further increase the mobility.

## IV. CONCLUSION

We synthesized $n$-type diamonds using TBP, which is much less toxic than phosphine. By incrementally increasing the gas flow rates of $H_2$ and $CH_4$, unintentional nitrogen incorporation was suppressed and the spin coherence time ($T_2$) increased to 1.62 ± 0.10 ms. From the ODMR spectra, we confirmed that all the measured NV centers were aligned along the [111] direction. The electrical properties of three of the synthesized phosphorus-doped diamond samples were evaluated using Hall measurements. They all exhibited $n$-type conduction, and the highest Hall mobility was estimated to be 422 $cm^2/(Vs)$ at room temperature. The realization of a long coherence time of perfectly aligned NV centers in $n$-type diamond prepared using TBP represents an important advancement toward enabling the manufacture of quantum diamond devices.

## ACKNOWLEDGEMENTS

This work was supported by OPERA (No. JPMJOP1841) and was supported in part by JSPS KAKENHI Grant No. 21H04653, MEXT Q-LEAP (JPMXS0118067395), the Collaborative Research Program of Institute for Chemical Research, Kyoto University (grant No. 2022-71, 2022-72), and the Spintronics Research Network of Japan.

## DATA AVAILABILITY

The data that support the findings of this study are available from the corresponding author upon reasonable request.

**Table 1.** Growth conditions for each sample. The samples are classified into four groups (O, A, B, and C) on the basis of the $H_2$ and $CH_4$ flow rates.

| Growth type | Sample name | Total gas flow rate (sccm) | $CH_4/H_2$ (%) | TBP/$CH_4$ (ppm) | Temp. (°C) | $C_N$ ($cm^{-3}$) | $C_P$ ($cm^{-3}$) | Longest $T_2$ (ms) |
|---|---|---|---|---|---|---|---|---|
| O | O1 | 100 | 0.1 | 26500 | 900 | - | - | 0.38 ± 0.03 |
| O | O1-S | 100 | 0.1 | 26500 | 900 | $1 \sim 2 \times 10^{17}$ | $2 \times 10^{16}$ | - |
| O | O2 | 100 | 0.1 | 26500 | 800 | - | - | 0.38 ± 0.05 |
| O | O2-S | 100 | 0.1 | 26500 | 800 | $0.8 \sim 1 \times 10^{17}$ | $4 \sim 6 \times 10^{16}$ | - |
| O | O3 | 100 | 0.1 | 26500 | 700 | - | - | 0.17 ± 0.02 |
| O | O3-S | 100 | 0.1 | 26500 | 700 | $7 \sim 8 \times 10^{16}$ | $3 \times 10^{16}$ | - |
| A | A1 | 100 | 1.0 | - | 1200 | - | - | 0.47 ± 0.03 |
| A | A2 | 100 | 1.0 | - | 1100 | - | - | 0.90 ± 0.06 |
| A | A2-S | 100 | 1.0 | - | 1100 | $5 \sim 8 \times 10^{16}$ | $2 \sim 3 \times 10^{17}$ | 0.59 ± 0.05 |
| A | A3 | 100 | 1.0 | - | 900 | - | - | 0.59 ± 0.10 |
| B | B1 | 400 | 0.1 | - | 800 | - | - | 1.38 ± 0.08 |
| B | B1-S | 400 | 0.1 | - | 800 | $2 \sim 3 \times 10^{16}$ | $1 \sim 4 \times 10^{16}$ | - |
| B | B2 | 400 | 0.1 | 6625 | 800 | - | - | - |
| B | B2-S | 400 | 0.1 | 6625 | 800 | $2 \sim 3 \times 10^{16}$ | $7 \times 10^{16}$ | - |
| B | B3 | 400 | 0.1 | 26500 | 800 | - | - | 0.28 ± 0.04 |
| B | B3-S | 400 | 0.1 | 26500 | 800 | $2 \times 10^{16}$ | $3 \sim 5 \times 10^{17}$ | - |
| C | C1 | 400 | 0.25 | - | 900 | - | - | 1.62 ± 0.10 |
| C | C2 | 400 | 0.25 | 1060 | 900 | - | - | 1.33 ± 0.05 |
| C | C2-S | 400 | 0.25 | 1060 | 900 | $1 \sim 2 \times 10^{16}$ | $2 \times 10^{17}$ | 1.01 ± 0.04 |
| C | C3 | 400 | 0.25 | 6625 | 800 | - | - | 0.68 ± 0.03 |

Table 2. List of parameters for each phosphorous-doped diamond film, as derived from Hall effect measurements.

| | $E_D$ [eV] | $N_D$ [cm$^{-3}$] | $N_A$ [cm$^{-3}$] | $\eta = N_A/N_D$ | $\mu$ [cm$^2$/Vs] |
|---|---|---|---|---|---|
| A2 | 0.62 | 7.0x10$^{16}$ | 4.5x10$^{16}$ | 64 % | 374 (340K) |
| B1 | 0.53 | 1.0x10$^{16}$ | 5.5x10$^{15}$ | 55 % | 422 (RT) |
| B2 | 0.53 | 4.5x10$^{16}$ | 2.5x10$^{16}$ | 55 % | 272 (RT) |

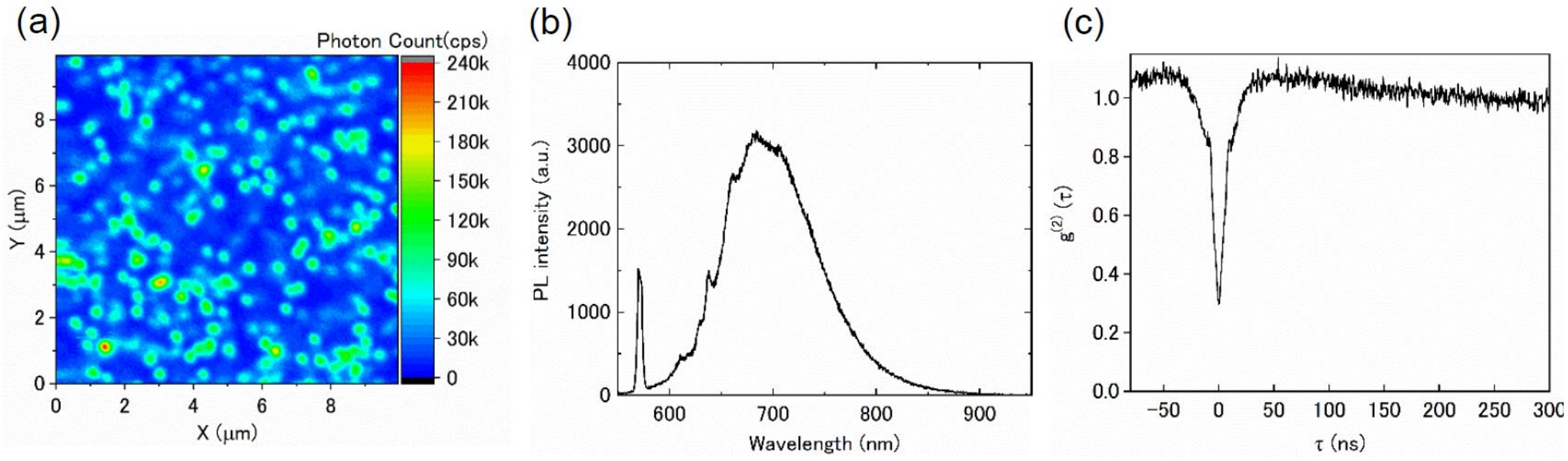

**Figure 1.** (a) Confocal fluorescence microscope image of a CVD film (from this picture, the NV concentration was estimated to be $10^{12}$ $cm^{-3}$). (b) A typical PL spectrum of an NV center. (c) Measurement of the second-order autocorrelation function $g^{(2)}(\tau)$ showing single-photon emission from a single NV center.

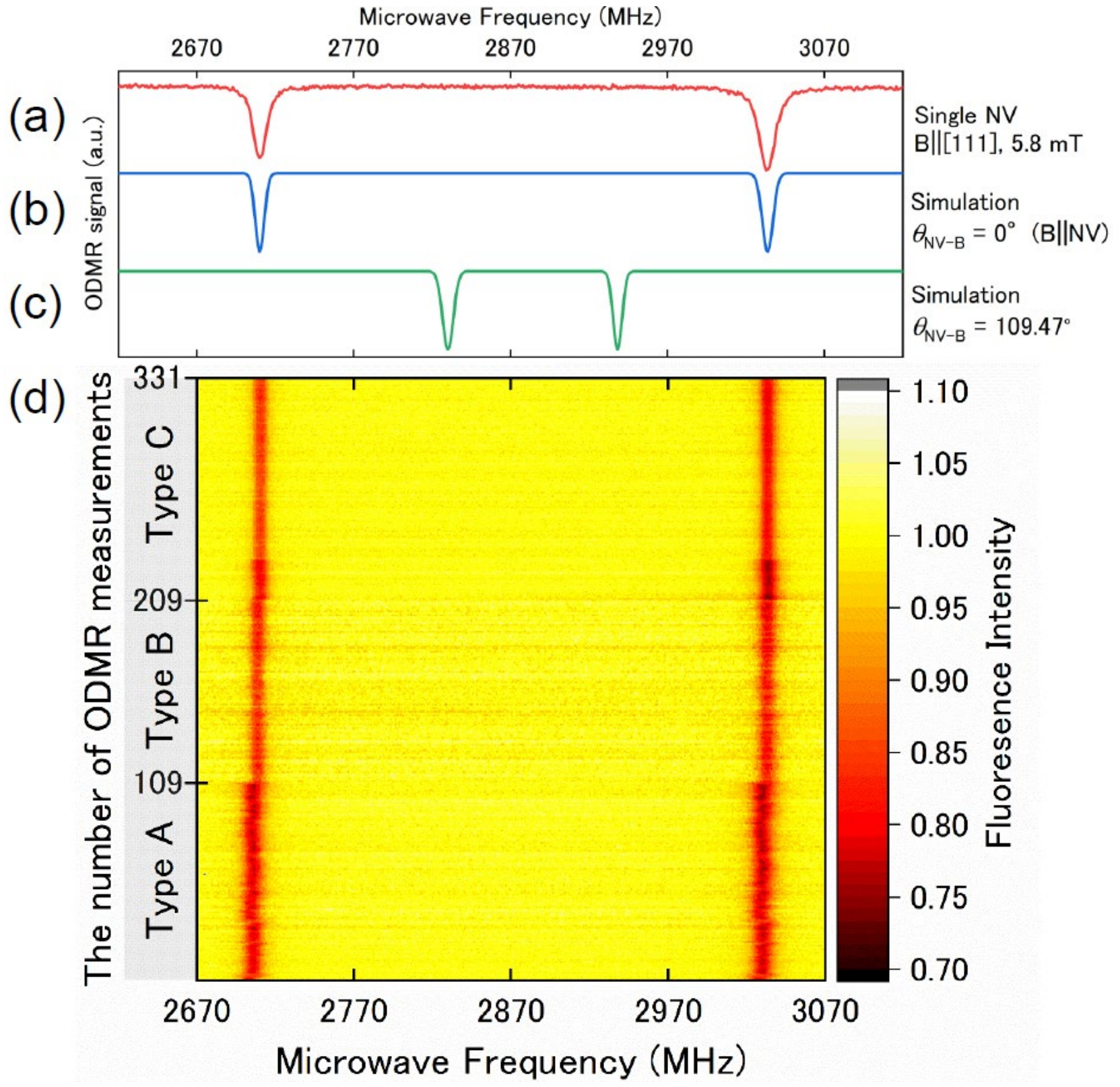

**Figure 2.** (a) ODMR spectrum obtained from a single NV center. The magnetic field (**B** ~ 5.8 mT) was applied along the [111] direction. (b) Simulated spectrum of a single NV center whose axis was aligned with the magnetic field (***B***||NV). (c) Simulated spectrum of a single NV center whose axis was inclined by 109.47° to the magnetic field ($\theta_{\text{NV-B}}$ = 109.47°). The direction corresponds to the $[1\bar{1}\bar{1}]$, $[\bar{1}1\bar{1}]$, or $[\bar{1}\bar{1}1]$ direction against the [111] direction. (d) Vertical stack of ODMR spectra of 331 different NV centers, which consisted of 109 NV centers in the type A sample, 100 NV centers in the type B sample and 122 NV centers in the type C sample. For each sample, NV centers were perfectly aligned in the [111] direction.

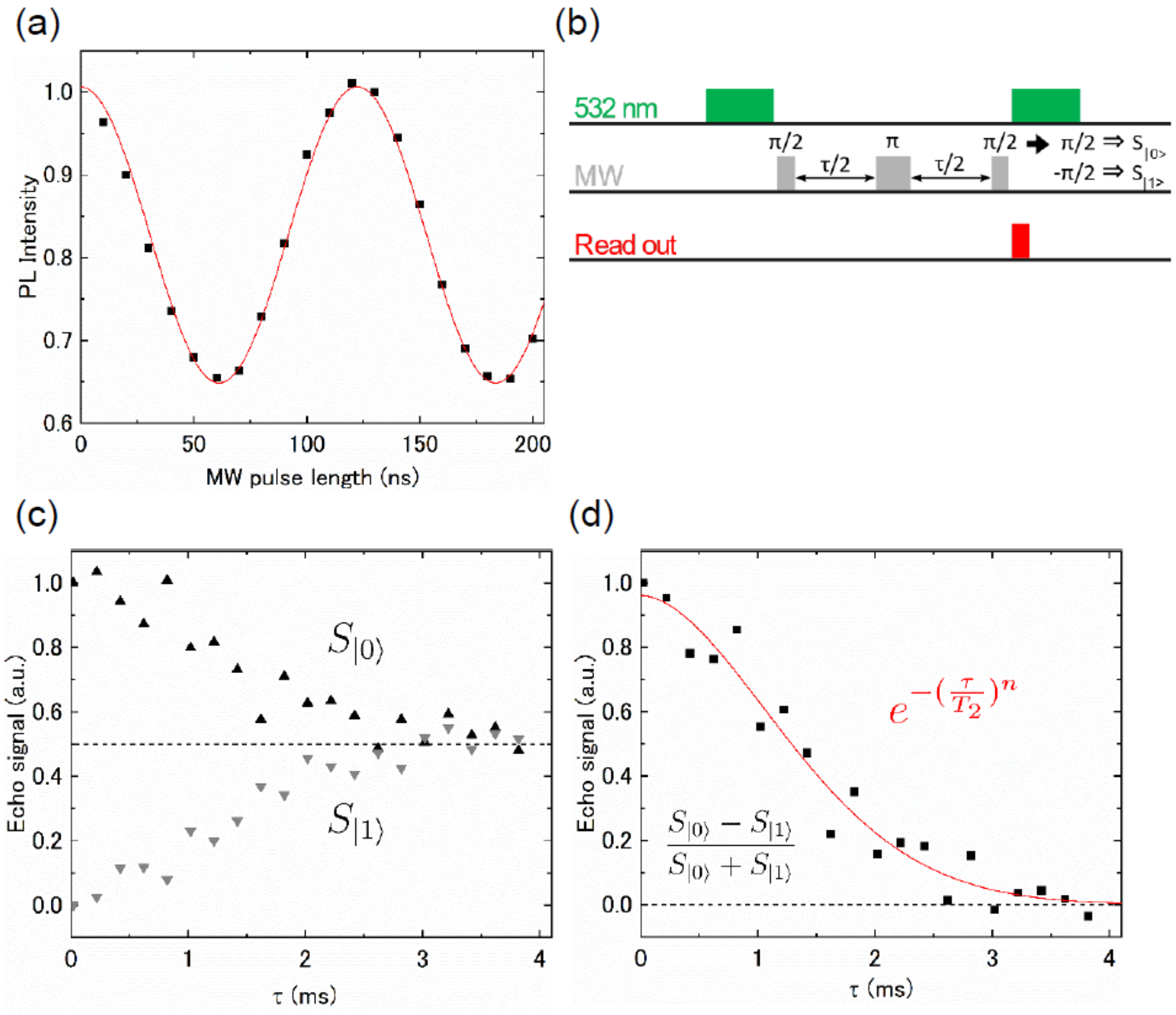

**Figure 3**. (a) Typical results of Rabi measurements (data with black dots, sinusoidal fit with red line, 36% contrast). (b) Pulse sequence for a Hahn echo measurement. When a final +π/2-pulse is applied, the spin is rotated toward the |0⟩ state (S|0⟩ measurement); for a final −π/2-pulse (or, alternatively, a 3/2-pulse), the spin is rotated toward the |1⟩ state (S|1⟩ measurement). (c) Results of the S|0⟩ measurement (black upward triangles) and of the S|1⟩ measurement (gray downward triangles). The middle black dashed line corresponds to both states being populated equally. (d) Echo signal derived from subtracting S|1⟩ from S|0⟩ (data with black dots, exponential-decay fit with red line, $T_2$ = 1.62 ± 0.1 ms). The dashed black line at 0 corresponds to the states being populated equally.

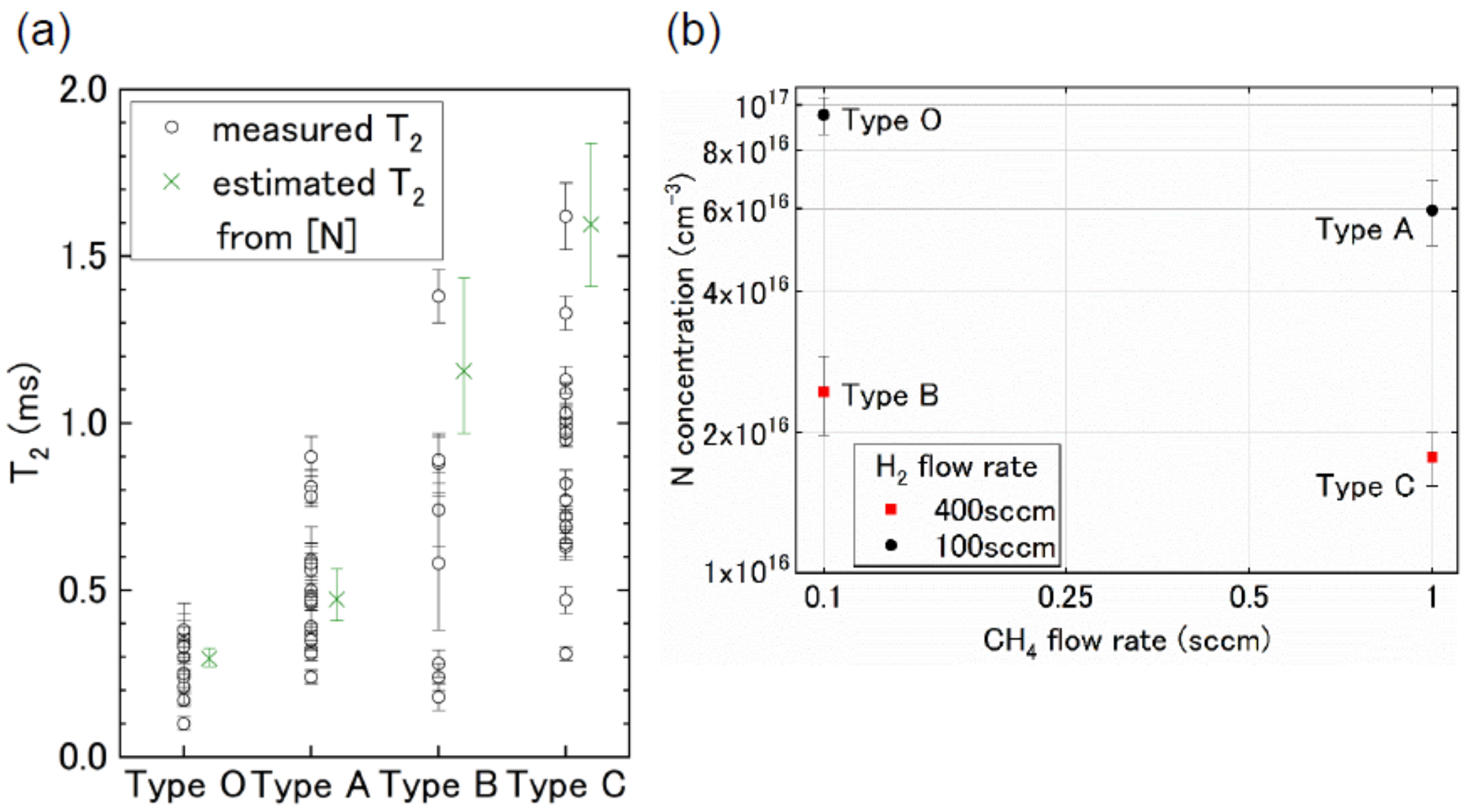

**Figure 4.** (a) $T_2$ for each growth type. The measured $T_2$ is shown as open circles. The longest $T_2$ for each growth type (O, A, B, and C) was 0.38 ± 0.03 ms, 0.90 ± 0.06 ms, 1.38 ± 0.08 ms, 1.62 ± 0.10 ms, respectively. The values of $T_2$ estimated from the nitrogen concentration of each growth type with a reported relationship 160 ± 12 μs ppm [3] are shown as green crosses to the right of the measured $T_2$. (b) The nitrogen concentration of each growth type is plotted against the $H_2$ and $CH_4$ flow rate during CVD growth. With decreasing nitrogen concentration, the $T_2$ increased.

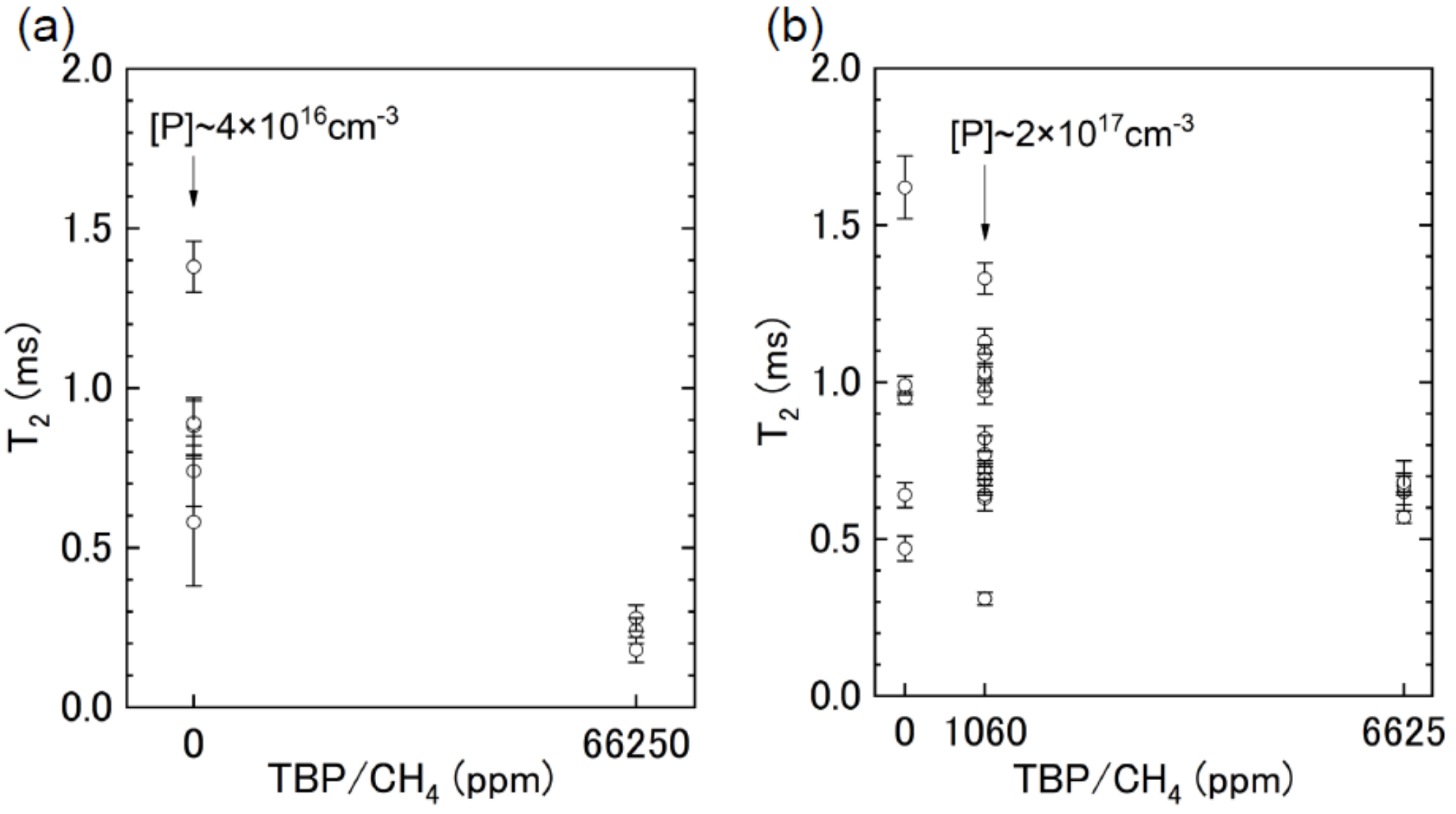

**Figure 5.** The $T_2$ plotted against different TBP flow rates is summarized for (a) growth-type B and (b) growth-type C. The SIMS measurements revealed that the phosphorus concentration is $4 \times 10^{16}$ cm$^{-3}$ with a TBP flow rate of 0 sccm for type B and $2 \times 10^{17}$ cm$^{-3}$ with a TBP flow rate of 1060 sccm for type C.

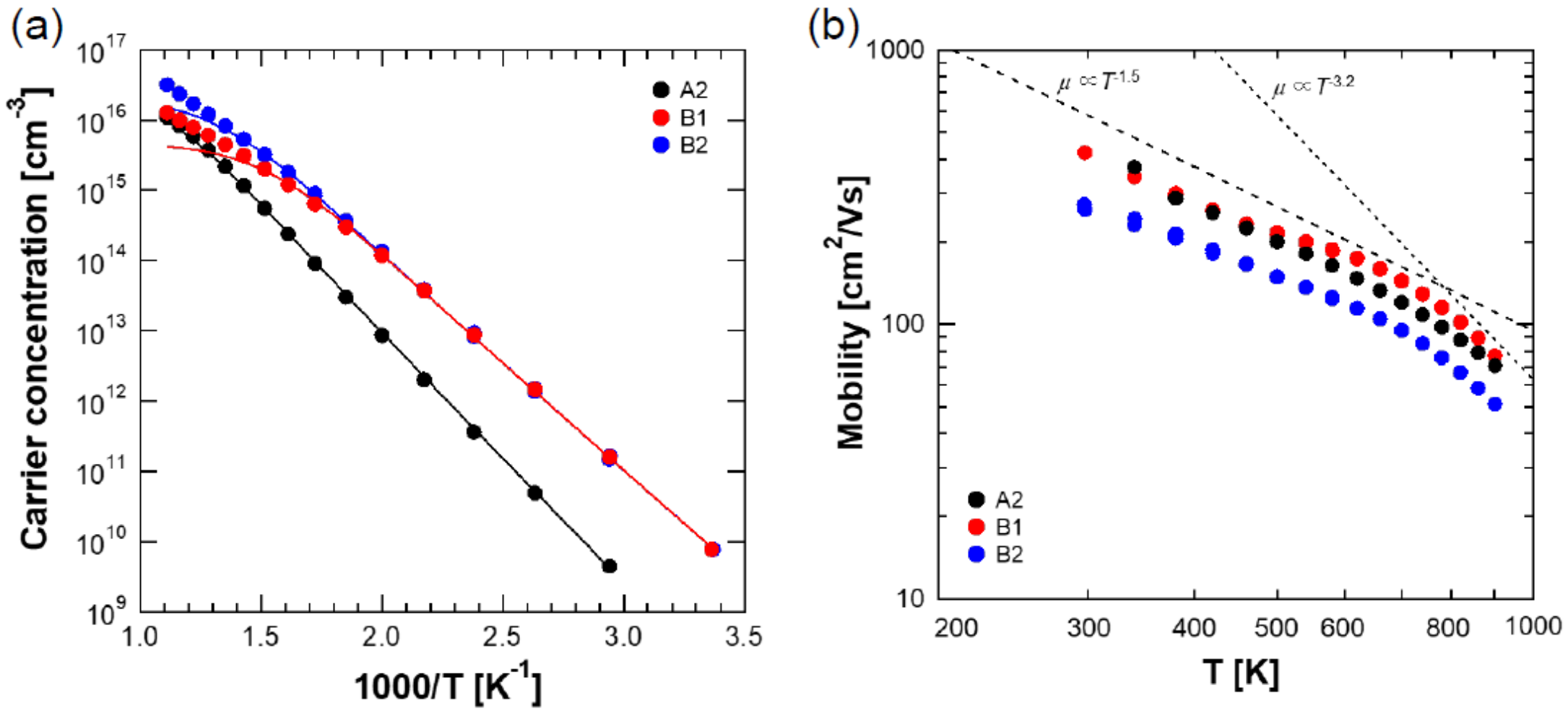

**Figure 6.** (a) Temperature dependence of the carrier concentration in phosphorus-doped diamond films, as determined by Hall effect measurements. The black circles represent sample A2 ($T_2$ ~ 0.90 ms), the red circles represent sample B1 ($T_2$ ~ 1.38 ms), and the blue circles represent sample B2. Each dashed line is a curve obtained by applying equation (1) to the measurement points and fitting using the least-squares approximation. (b) Temperature dependence of electron mobility for phosphorus-doped diamond films. The black circles represent sample A2 ($T_2$ ~ 0.90 ms), the red circles represent sample B1 ($T_2$ ~ 1.38 ms), and the blue circles represent sample B2. The dashed line visually shows that the temperature dependence of electron mobility follows acoustic phonon scattering ($\mu \propto T^{-1.5}$) and other factors ($\mu \propto T^{-3.2}$).